\documentclass[10pt,a4paper,twocolumn]{article}

\usepackage{graphicx}
\usepackage{dcolumn}
\usepackage{bm}
\usepackage{textcomp}
\usepackage{color}
\usepackage[normalem]{ulem}
\usepackage{gensymb}
\usepackage{amsmath}
\usepackage{acronym}
\usepackage{authblk}

\definecolor{JLAQ}{rgb}{1,0,1}

\begin{document}
\bibliographystyle{ieeetr}
\title{Broadband Huygens' Metasurface Based on Hybrid Resonances}
\author[1]{M. Londo\~no\thanks{malondonoc@unal.edu.co}}
\author[2]{A. Sayanskiy\thanks{a.sayanskiy@metalab.ifmo.ru}}
\author[3]{J. L. Araque-Quijano\thanks{jlaraqueqd@unal.edu.co}}
\author[2]{S. B. Glybovski\thanks{s.glybovski@metalab.ifmo.ru}}
\author[1]{J. D. Baena\thanks{jdbaenad@unal.edu.co}}

\affil[1]{Physics Department, Universidad Nacional de Colombia, Bogota 111321, Colombia}
\affil[2]{Department of Nanophotonics  and  Metamaterials,  ITMO  University,  St.  Petersburg 197101, Russia}
\affil[3]{Electrical Engineering Department, Universidad Nacional de Colombia, Bogota 111321, Colombia}

\maketitle

\section{Abstract}
Recently, a special class of Huygens' surfaces has been proposed which are capable of manipulation of transmitted wavefronts while exhibiting high transparency over a broad range of frequencies. In this work we propose and study a new meta-atom for achieving broadband transparency of Huygens' surfaces based on two mutually shifted split-ring resonators. The resonators were carefully optimized using a new developed analytical model so that their electric and magnetic responses was balanced to achieve unidirectional scattering of the incident field. Moreover, their mutual shift canceled their inductive coupling so that both the responses were resonant at the same frequency and the meta-atom exhibited no magneto-electric coupling. The analytical prediction of broadband transparency along with the frequency-dependent phase shift of the transmitted wave within the full coverage of $360\degree$ has been verified numerically and experimentally in the microwave range. The proposed metasurface can be used for symmetric high-efficient polarization conversion and beam-splitting, and opens the way for other interesting applications.


\section{Introduction}

Over the past decade, research on electromagnetic metasurfaces (MSs) has quickly flourished and brought many applications ranging from microwaves to optics. In general, MSs are conceived as 2D arrays of electromagnetic scatterers whose size and separation are much smaller than the operational wavelength. They can be approximated as a combination of smooth electric and magnetic surface current densities representing the averaged responses of discrete unit cells respect to an incident wave. Properties of a specific MS come from its unit-cell microstructure, which determines both the induced surface currents and, as a result, the scattered field. By properly designing the unit cell a lot of functionalities have been proposed for MSs during the last years, ranging from Frequency Selective Surface (FSS) to dynamically tunable and reflectionless phase holograms. In order to appreciate the wide range of realizations and functions, one can refer to a number of review papers \cite{Holloway12,Capasso14,Minovich2015195,Chen2016,Estakhri2016A21,Glybovski2016}.

There is a special class of reflectionless MSs, which are capable of changing the magnitude and/or phase of selected components of the incident wave causing no reflection. In present literature they use to be called Huygens' MSs. For such MSs the effective electric and magnetic surface currents induced simultaneously must be balanced (i.e. they must scatter equally strong waves). It is achieved by using Huygens' sources, possessing simultaneously induced electric and magnetic dipoles which are orthogonal one to each other, equally strong, and excited in phase. Inside this category one can distinguish two extreme possibilities: perfect absorption or perfect transparency \cite{Kim2014,Kwon2018}. The scope of this paper is focused on the last type: perfect transparency. This type of Huygens' MSs were first experimentally demonstrated using combinations of dipole and loop printed particles in the microwave frequency band \cite{Pfeiffer2013} and Mie-type resonances of silicon nanodisks in optics \cite{Decker2015}.

By slowly varying structural parameters of unit cells over the MS, one is able to reproduce the distribution of both the averaged surface currents to support practically any desired transmitted fields with illumination by a given field source still with no reflection \cite{Epstein20145680}. Following this approach, reflectionless beam shaping (refraction, focusing etc.) has been demonstrated using Huygens' surfaces in the microwave \cite{Pfeiffer2013, Wong2014360, Wong2015913, Jia2015, Wong20161293, Wan20174427}, the millimeter-wave \cite{Pfeiffer20134407}, and the optical \cite{Alu2013,Pfeiffer20142491, Gupta20161641, Zhao2016181102} ranges. Other functions of Huygens' surfaces found in literature are: creating optical vortex beams (also known as orbital angular momentum states) \cite{Chong20155369}, beam splitting \cite{Kim2014,Jia2015545}, controlling antenna radiation patterns \cite{Mehdipour2015978,Chen2015,Epstein2016,Zhengbin201710} and holography \cite{Chong2016514, Zhao2016, Huang2017}.
Moreover, dynamically tunable Huygens' surfaces have been recently reported \cite{Chen2017}.

In most cases, MSs are reflectionless only in a relatively narrow frequency range, where its wave control functions are observed, and become highly reflective at other frequencies, which limits possible applications. Therefore, special attention has been devoted in the last few years to broadband reflectionless MSs, first proposed in \cite{Asadchy_PRX_2015}. Here it was first shown that, even though the perfect absorption bandwidth of thin MSs is fundamentally limited, there is no bandwidth limitation for the reflectionless behavior. Consequently, the narrow band nature of the field manipulation function of a single MS of this type does not limit applications since the broadband property of low reflection allows one to cascade such MSs providing multiple functionalities at separate frequencies \cite{Elsakka2016}. Apart from perfect absorption \cite{Asadchy_PRX_2015,Cole2016}, broadband reflectionless MSs have been designed to provide such frequency-selective functions as beam refraction and shaping \cite{Elsakka2016} and twist-polarization \cite{Faniayeu2017}. As shown in \cite{Asadchy_PRX_2015}, constructing a broadband reflectionless MS requires the electric and magnetic response to be resonant and balanced within all the resonance band of unit  cells. For that, it was proposed to use the one and the same resonant mode of chiral helix meta-atoms creating both the electric and magnetic responses, where magnetoelectric coupling was compensated by racemic arrangement of the helices with opposite handedness.   

\begin{figure}
  \centering
  \includegraphics[width=0.9\columnwidth]{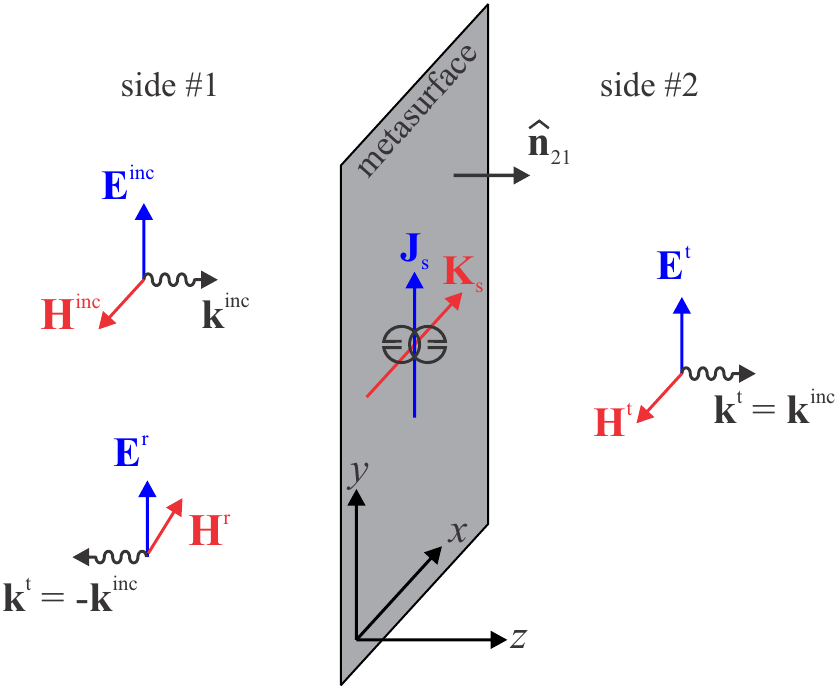} 
  \caption{Diagram of a metasurface illuminated by an incident plane wave ($\textbf{E}^\text{inc}$, $\textbf{H}^\text{inc}$). Transmitted ($\textbf{E}^\text{t}$, $\textbf{H}^\text{t}$) and reflected ($\textbf{E}^\text{r}$, $\textbf{H}^\text{r}$) plane waves are shown. One unit cell is represented at the center. $\textbf{J}_\text{s}$ and $\textbf{K}_\text{s}$ represents the effective electric and magnetic surface current densities, respectively.}
  \label{Fig_metasurface}
\end{figure}

In this work we design a new broadband reflectionless MS providing any phase shift based on a new type of unit cell, shown in Fig.~\ref{Fig_metasurface} and Fig.~\ref{Fig_few_cells}(a). The unit cell is symmetric and composed of two Split Ring Resonators (SRRs) lying on two parallel planes separated by a small distance $d$ and being laterally shifted. For a certain lateral shift of the SRRs their mutual inductance is canceled \cite{Roemer1990}, so that the introduced particle possesses degenerate electric and magnetic resonant modes with no magnetoelectric coupling. As it will be shown, as both the modes resonate at the same frequency and have similar dispersion around the resonance, the transmission coefficient magnitude holds close to 1 in a wide frequency band, while simultaneously its phase changes with frequency in a full $2\pi$ range.

This paper is organized as follows. In Section \ref{Theory} we analytically model the behavior of a metasurface composed of unit cells consisting of a pair of mutually shifted SRRs. Section~\ref{Simulation} is devoted to present numerical results for an idealized broadband Huygens' MS composed of lossless and substrateless SRRs, including the study of a finite sample of MS. In Section~\ref{Experiment} the experimental results are given to support the theoretical predictions. Section~\ref{Applications} summarizes some of the applications envisioned for the MS proposed.

\section{Theory}\label{Theory}

\subsection{Reflectionless surface condition}

Let us consider the situation of Fig.~\ref{Fig_metasurface}, where a homogeneizable metasurface composed of electrically and magnetically polarizable meta-atoms is illuminated by an incident plane wave propagating along the $z$-axis and linearly polarized along the $y$-axis. For simplicity, let us consider meta-atoms are periodically distributed over a square net with periodicity $a$. The effective electric and magnetic surface current densities can be calculated through the corresponding induced dipoles $p_y$ and $m_x$ of one meta-atom as $J_{\text{s}y}=-\text{i}\omega p_y/a^{2}$ and $K_{\text{s}x}=-\text{i}\omega\mu_{0}m_{x}/a^{2}$. 
Any magneto-electric coupling is absent in this model since the meta-atom is symmetric under inversion. Inserting the surface currents into the boundary conditions $\hat{\textbf{n}}_{21}\times(\textbf{H}_2-\textbf{H}_1)=\textbf{J}_\text{s}$ and $\hat{\textbf{n}}_{21}\times(\textbf{E}_2-\textbf{E}_1)=-\textbf{K}_\text{s}$, where $\hat{\textbf{n}}_{21}$ is the unitary normal vector pointing from side \#1 to side \#2, and taking into account the general plane wave relation $\textbf{H}=Z_0^{-1}\hat{\textbf{k}}\times\textbf{E}$, where $\hat{\textbf{k}}$ is an unitary vector parallel to the wavevector and $Z_0$ is the free space wave impedance, we obtain the following equations:
\begin{equation}
 \label{boundaryConditions}
 \begin{gathered}
  t_{y}-1+r_{y} = \frac{\text{i}k}{a^{2}} \frac{p_y}{\epsilon_0 E^\text{inc}_y} = \frac{\text{i}k}{a^{2}} \alpha^\text{e}_{y}   \\
  t_{y}-1-r_{y} = \frac{\text{i}k}{a^{2}} \frac{m_{x}}{H^\text{inc}_x} = \frac{\text{i}k}{a^{2}} \alpha^\text{m}_{x}
 \end{gathered}
\end{equation}
where $t_{y}$ and $r_{y}$ are the transmission and reflection coefficients for $y$-polarized waves, respectively, $k=\omega\sqrt{\mu_0\epsilon_0}$ is the wavenumber, and $\alpha^\text{e}_{y} = p_y/(\epsilon_0 E^\text{inc}_y)$ and $\alpha^\text{m}_{x} = m_{x}/H^\text{inc}_x$ are the effective electric and magnetic polarizabilities of the unit cell along the $y$-axis. It is clear from (\ref{boundaryConditions}) that, to get total transmission with an arbitrary phase shift, i.e. $r_y=0$ and $t_y=\text{exp}(\text{i}\phi)$, we must force
\begin{equation}
\label{reflectionless_condition}
\alpha^\text{e}_{y} = \alpha^\text{m}_{x}=\text{i}(1-\text{exp}(\text{i}\phi))a^2/k, 
\end{equation}
where $\phi$ the phase shift provided by the MS.

\begin{figure}
  \centering
  \includegraphics[width=0.9\columnwidth]{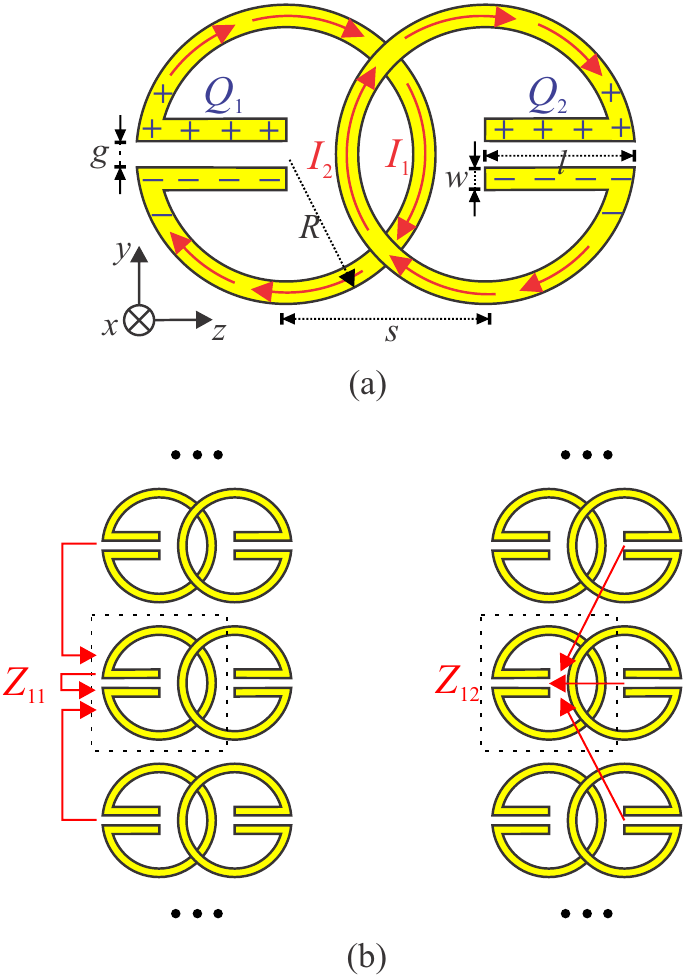} 
  \caption{a) The unit element of the metasurface is formed by two separated SRRs placed on two close parallel planes. The conventions for positive current and positive charge are indicated. b) Illustration of the meaning of the impedances $Z_{11}$ and $Z_{12}$ used in (\ref{circuitEquations}).}
  \label{Fig_few_cells}
\end{figure}

\subsection{Equivalent circuit model and polarizabilities}

Before solving the coefficients $t_y$ and $r_y$ from (\ref{boundaryConditions}), we need to calculate the polarizabilities of an unit cell of the MS. Let us consider the unit cell containing the meta-atom shown in Fig.~\ref{Fig_few_cells}(a) with the conventions therein for currents along the rings $I_1$ and $I_2$, and charges for the legs, $Q_1$ and $Q_2$. The electric and magnetic dipoles can be calculated as:
\begin{equation}
 \label{dipoles}
 \begin{gathered}
  p_y = d(Q_1+Q_2) = \frac{d}{\text{i}\omega}(I_1-I_2) \\
  m_x = A(I_1+I_2)
 \end{gathered}
\end{equation}
where $I_1$ and $I_2$ are arbitrarily defined as the current at the center of each metallic strip, $d$ is the effective distance between the positive and negative centers of charge, and $A$ is the effective area of a loop of uniform current holding the same magnetic dipole as one single SRR. By assuming a cosine profile for the current along the strips with nodes at the ends, both effective geometric parameters, $d$ and $A$, can be approximated in terms of the actual external radius of the ring $R$ and the length of the leg $l$ of each SRR as (see the Appendix):
\begin{equation}
 \label{effGeoPar}
  d \approx 2R \frac{sin(\pi \xi^{-1})}{\xi-\xi^{-1}}, \ \ \ \ \
  A \approx \pi R^2 \frac{sin(\pi \xi^{-1})}{\pi \xi^{-1}}
\end{equation}
where $\xi=2(1+l/\pi R)$. We can approximate the values of the currents by solving the next two coupled circuit equations:
\begin{equation}
 \label{circuitEquations}
 \begin{gathered}
  Z_{11} I_{1} + Z_{12} I_{2} = 
  \text{i}\omega A \mu_{0}H_x^{\text{inc}}-d E_y^{\text{inc}} \\
  Z_{21} I_{1} + Z_{22} I_{2} = 
  \text{i}\omega A \mu_{0}H_x^{\text{inc}}+d E_y^{\text{inc}}
  \end{gathered}
\end{equation}
Here $Z_{11}$ is the self-impedance of one ring plus the mutual impedances with all the rings of the same side \#1 of the MS, and $Z_{12}$ is the mutual impedance between one ring on one side \#1 and all the rings on the opposite side \#2 (see Fig.~\ref{Fig_few_cells}(b)). In other words, the impedances $Z_{11}$ and $Z_{12}$ take into account mutual coupling of meta-atoms in the 2D array. By symmetry, it is clear that for the rings on the opposite side \#2 one can write the same values: $Z_{22}=Z_{11}$ and $Z_{21}=Z_{12}$. The terms on the right side of (\ref{circuitEquations}) represent the electromotive forces coming from the power applied by the incident fields over the dipoles. Solving (\ref{circuitEquations}) the currents are 
\begin{equation}
 \label{currents}
  I_{1/2} = \frac{\text{i}\omega A \mu_{0}H_x^{\text{inc}}}{Z_{11}+Z_{12}}
          \mp \frac{d E_y^{\text{inc}}}{Z_{11}-Z_{12}}
\end{equation}
From this formula, it is clear that the magnetic field contributes to the excitation of the even mode with $I_1=I_2$ while the electric field excites the odd mode with $I_1=-I_2$. In the case of plane wave incidence both fields are simultaneously acting on the meta-atom, so that in general both the even and the odd modes are excited in a certain combination. Now, by substituting these currents into (\ref{dipoles}), the polarizabilities of the meta-atom are obtained:
\begin{equation}
 \label{polarizabilities}
 \begin{gathered}
  \alpha^\text{e}_y = \frac{p_{y}}{\epsilon_0 E_y^{\text{inc}}}=-\frac{2d^{2}}{\text{i}\omega\epsilon_0 (Z_{11}-Z_{12})} \\
  \alpha^\text{m}_x = \frac{m_{x}}{H_x^{\text{inc}}}=\frac{2\text{i}\omega\mu_0 A^2}{Z_{11}+Z_{12}}
 \end{gathered}
\end{equation}
From (\ref{currents}) and (\ref{polarizabilities}), currents and polarizabilities may resonate if the reactive parts of the impedances cancel out.  If $\Im(Z_{11} + Z_{12})=0$, then the currents are more even than odd and thus the resonant mode is predominantly magnetic. On the contrary, if $\Im(Z_{11} - Z_{12})=0$, then the currents are more odd than even and then the resonance is mainly electric. Most likely the two resonant modes happen at different frequencies, unless $\Im(Z_{12})=0$. In this very special case the magnetic and electric resonances are degenerate and the meta-atom possesses both the electric and magnetic responses. It is important to mention that, however, there is no magnetoelectric coupling due to the inversion symmetry of the meta-atom with respect to its geometrical center. The degenerate resonance condition $\Im(Z_{12})=0$ is possible as long as the total mutual inductance of two partially overlapped current loops may be made equal to zero \cite{Roemer1990}. In the proposed geometry it is realized by the choice of the lateral shift $s$ of the SRRs (see Fig.~\ref{Fig_few_cells}(a)) positioned in slightly different parallel planes.

\subsection{Transmission and reflection coefficients}

Now, inserting (\ref{polarizabilities}) into (\ref{boundaryConditions}) and solving the system of equations, we get the transmission and reflection coefficients of the MS:
\begin{equation}
 \label{coefficients}
 \begin{gathered}
  t_y = 1-\frac{Z_{0}}{a^{2}}\left[\frac{k^2 A^2}{Z_{11}+Z_{12}} + \frac{d^{2}}{Z_{11}-Z_{12}} \right] \\
  r_y = \frac{Z_{0}}{a^{2}}\left[\frac{k^2 A^2}{Z_{11}+Z_{12}} - \frac{d^{2}}{Z_{11}-Z_{12}} \right]
 \end{gathered}
\end{equation}
In the first order of approximation, $Z_{11} = R_{11} - \text{i}\omega L + \text{i}/(\omega C)$ and $Z_{12} = R_{12}-\text{i}\omega M$, where the resistances $R_{11}$ and $R_{12}$ may be related with the dissipated power and scattered power, $L$ is the magnetic inductance of one ring, $C$ is the electric capacitance between the legs of the SRR, and $M$ is the mutual inductance between the two shifted SRRs in one unit cell. To be accurate, all these quantities may include the effect of the neighboring SRRs in the same side of the MS, as sketched in Fig.~\ref{Fig_few_cells}(b). If there was no energy dissipation (i.e. perfectly conducting SRRs in free space) $\left|t_y\right|^2+\left|r_y\right|^2=1$ and, using (\ref{coefficients}), it is a straightforward task to obtain the following formulas for the scattering resistances:
\begin{equation}
 \label{resistances}
   R_{11/12} = Z_0 \frac{k^2 A^2 \pm d^2}{2a^2}
\end{equation}
where the sign in the numerator is ``+" for $R_{11}$ while ``-" for $R_{12}$.

In order to avoid any reflection, it is clear from (\ref{coefficients}) that the circuits should be decoupled with $Z_{12}=0$ and, at the same time, the geometrical condition $d=kA$ should be satisfied. In fact, based on (\ref{resistances}), the last condition means also that $R_{12}=0$. Regarding the reactive part of $Z_{12}$, since $M$ is positive for coaxial rings while negative for coplanar rings, it can be canceled out by carefully adjusting the value of $s$ \cite{Roemer1990}. At least this approximation is valid for SRRs much smaller than a wavelength.
However, the condition $d=kA$ cannot be exactly met at all frequencies. In our design we should force it only about the resonance frequency, i.e. in the frequency range where interaction of the incident wave with the MS becomes strong. In contrast, outside of the resonance band, the structure it almost transparent for the incident wave, so that the last condition becomes unimportant. It means that the geometry should actually satisfy $d=k_0 A$, where $k_0$ is the wavenumber at the resonance frequency. Using the formulas for effective distance and effective area in (\ref{effGeoPar}), a relation for the optimal length of the leg can be obtained:
\begin{equation}
 \label{balanceCondition}
 l=\pi R \left( -1+\frac{1}{2} \sqrt{1+ \frac{\lambda_0}{\pi R}} \right)
\end{equation}
where $\lambda_0$ is the free space wavelength at the resonant frequency. Under all the mentioned conditions, i.e. assuming a perfectly balanced geometry for the unit cell, the transmission and reflection coefficients can be approximated as follows:
\begin{equation}
 \label{simplifiedCoefficients}
 t_y \approx 1-\frac{2Z_0(d/a)^2}{Z_0(d/a)^2 - \text{i}\omega L + \frac{\text{i}}{\omega C}}, \ \
 r_y \approx 0
\end{equation}
It is a straightforward task to confirm from (\ref{simplifiedCoefficients}) that $\left|t_y\right|=1$, at the resonance frequency, meaning that the incident power is totally transmitted through the MS. Moreover, the resonant behavior of $t_y$ guarantees that its phase covers the full range between $0\degree$ and $360\degree$.

\section{Numerical validation}\label{Simulation}

\begin{figure}
  \centering
  \includegraphics[width=0.9\columnwidth]{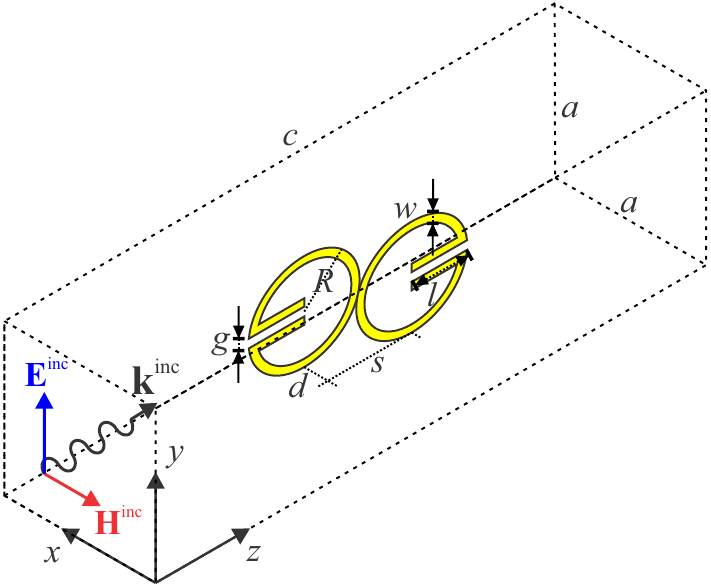} 
  \caption{The simulated unit cell where $x$- and $y$-walls are periodic boundaries while Bloch-Floquet ports are placed at $z$-walls. The incident wave is traveling along the z-axis. The geometrical parameters are (mm): $a=4$, $c=100$, $R=1.7$, $w=0.05$, $g=0.05$, $l=3$, and $d=0.5$. The lateral shift $s$ was swept from 0 to 4~mm, being the optimal value $s=2.4$~mm.}
  \label{Fig_sim_box}
\end{figure}

To confirm the above analytical predictions, the MS composed of lossless meta-atoms depicted in Fig.~\ref{Fig_sim_box} with no supporting material was simulated using the commercial software CST Microwave Studio in the frequency range 0-10~GHz for the infinite case, and using an Electromagnetic simulation code developed at Universidad Nacional de Colombia to study the effects resulting from the finite extent of real MSs at resonance.

\subsection{Simulated transmission and reflection}

\begin{figure}
  \centering
  \includegraphics[width=0.9\columnwidth]{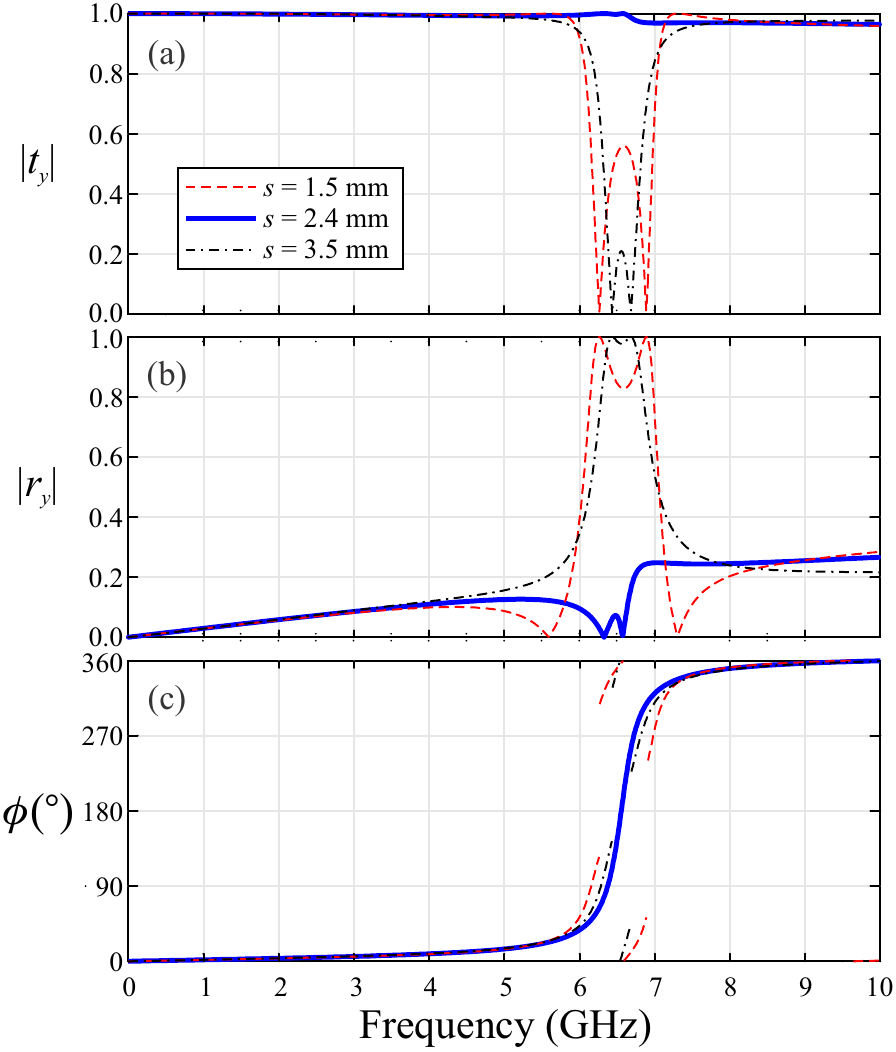} 
  \caption{Simulated transmission and reflection coefficients for the geometry of Fig.~\ref{Fig_sim_box}. Magnitudes of $t_y$ and $r_r$ are shown in (a) and (b) while only the phase of $t_y$ is depicted in (c).}
  \label{Fig_sim_trans_refl}
\end{figure}

Simulated transmission and reflection coefficients for a $y$-polarized incident wave are shown in Fig.~\ref{Fig_sim_trans_refl} for different values of the shift parameter $s$. In the case of $x$-polarized incident waves, the response is trivially $t_x = 1$, because the incident electric field is orthogonal to the plane of the SRRs and does not induce any currents. However, for the case of $y$-polarized incident waves, we note that the resonance is split into two well-separated frequencies for most curves due to the coupling between the two rings in the unit cell. It will be demonstrated below that one of the two dips corresponds to a magnetic resonance or even mode ($I_1=I_2$) while the other one is connected with the electric or odd mode ($I_1=-I_2$). There exists an interval $2~\text{mm}<s<3~\text{mm}$ for which the resonance split is not observable, because this split is smaller than the bandwidth of the resonance. Interestingly, there is an exact value of shift $s=2.4~\text{mm}$ for which the mutual coupling between the rings cancels out exactly and, consequently, the resonance is not split (see the thick blue line in Fig.~\ref{Fig_sim_trans_refl}(a)). This corresponds to the particular case where the even and odd modes resonate at the same frequency. In fact, for this value of $s$, the reflection magnitude is lower than 0.3 for the whole considered range of frequencies (see the thick blue line in Fig.~\ref{Fig_sim_trans_refl}(b)), i.e. less than 9\% of the incident power is reflected. Furthermore, the $y$-polarized waves undergoes almost full transmission with a frequency-dependent phase shift (see the thick blue line in Fig.~\ref{Fig_sim_trans_refl}(c)). Furthermore, from Fig.~\ref{Fig_sim_trans_refl}(c), it is clear that any phase shift is available in transmission for $y$-polarized waves from the full range between 0 and $360\degree$. Unlike the magnitude plot, the phase plot clearly shows the position of the resonance frequency at about 6.6~GHz, where the phase take the exact value of $180\degree$. It is important to mention that, by introducing the corresponding resonance wavelength $\lambda_0=45$~mm and the used radius $R=1.7$~mm into (\ref{balanceCondition}), the expected leg length $l$ should be about 2.9~mm which is very close to the optimal value of $l=3$~mm obtained by parameter sweeping.

\subsection{The hybrid electric/magnetic resonance}

\begin{figure}
  \centering
  \includegraphics[width=0.9\columnwidth]{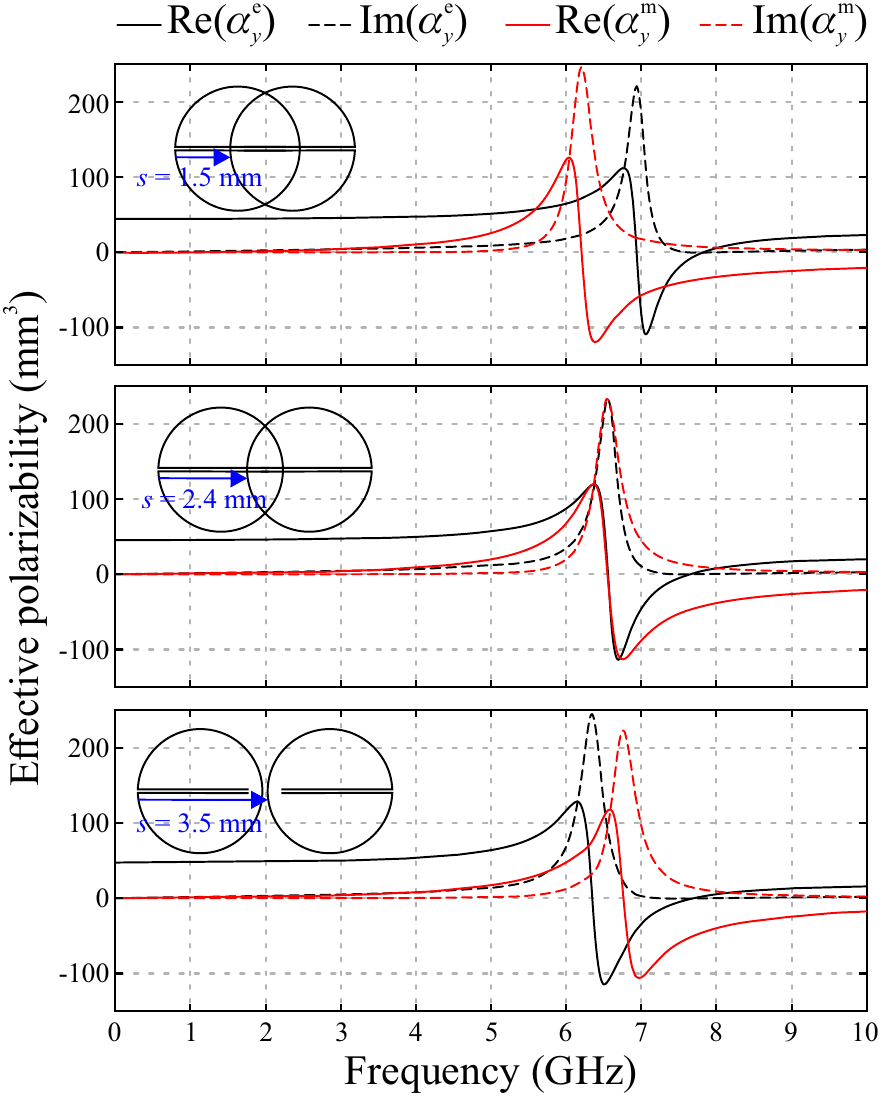} 
  \caption{Electric and magnetic polarizabilities retrieved from simulated reflection and transmission coefficients using (\ref{boundaryConditions}).}
  \label{Fig_sim_polarizabilities}
\end{figure}

\begin{figure}
  \centering
  \includegraphics[width=0.9\columnwidth]{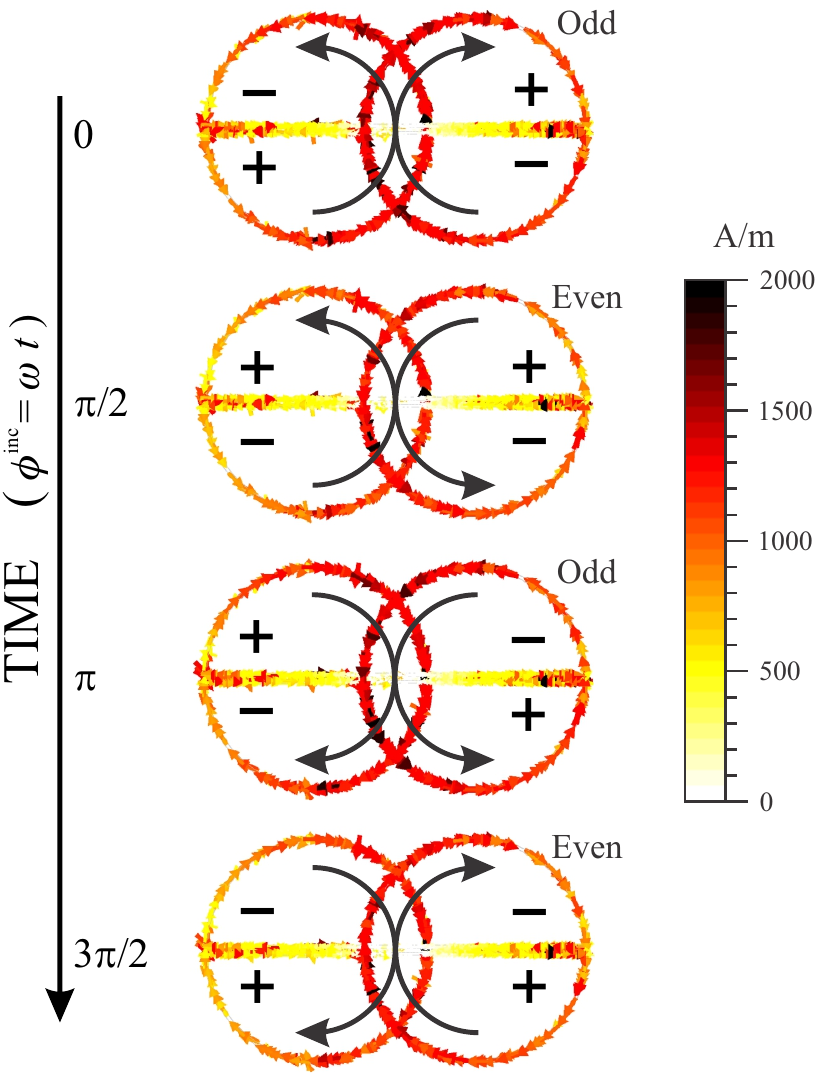} 
  \caption{Surface current distribution of the balanced hybrid electric/magnetic resonant mode at 6.6~GHz for the geometrical parameters of Fig.~\ref{Fig_sim_box} with $s=2.4$~mm.}
  \label{Fig_hybrid_mode}
\end{figure}

In order to demonstrate the nature of each resonance, polarizabilities were retrieved from the numerically calculated transmission and reflection coefficients by using (\ref{boundaryConditions}) and are depicted in Fig.~\ref{Fig_sim_polarizabilities}. As seen from the simulation results, in the case of $s=1.5$~mm ($s=3.5$~mm) the first resonance is mainly magnetic (electric) while the second one is mainly electric (magnetic). Both resonances just coincide when the lateral shift is $s=2.4$~mm. In fact, this is the desired stage of the mentioned degenerate resonance. To illustrate how the unit cell operates when excited by a plane wave, the surface currents at different time moments within a whole oscillation period are shown in Fig.~\ref{Fig_hybrid_mode}. It is worth noting that this resonant mode is the linear combination of the even and odd modes in the sense that half of the time the current pattern looks like an even mode, while in the other half it looks like an odd mode.

\subsection{Numerical study of a finite metasurface}\label{FiniteSimulation}

As a further numerical validation, the simulation of a finite MS has been carried out to better assess the experimental results affected by the practical limitations imposed by the manufacture and measurement processes. For this purpose, the Advanced Metal-Dielectric Solver (AMDS) code, developed at Universidad Nacional de Colombia was employed, which is a full-wave simulator based on the Method of Moments (MoM) technique \cite{Harrington1993} with Multi-Level Fast Multipole Method (ML-FMM) \cite{Coifman1993} and optimizations specific for finite-periodic structures such as the one under analysis. This code has been validated with well-known reference results, among which we could cite \cite{Peña2009} with a freely available code to compute scattering by layered dielectric spheres. This MoM implementation uses the Rao-Wilton-Glisson (RWG) basis \cite{Wilton1982}, which is defined on pairs of cells of meshes composed of triangular facets. Each basis element provides a piecewise linear approximation to surface currents and thanks to the div-conforming property, the overall representation correctly accounts for both charge and current contributions to the resulting fields.

\begin{figure}
  \centering
  \includegraphics[width=0.9\columnwidth]{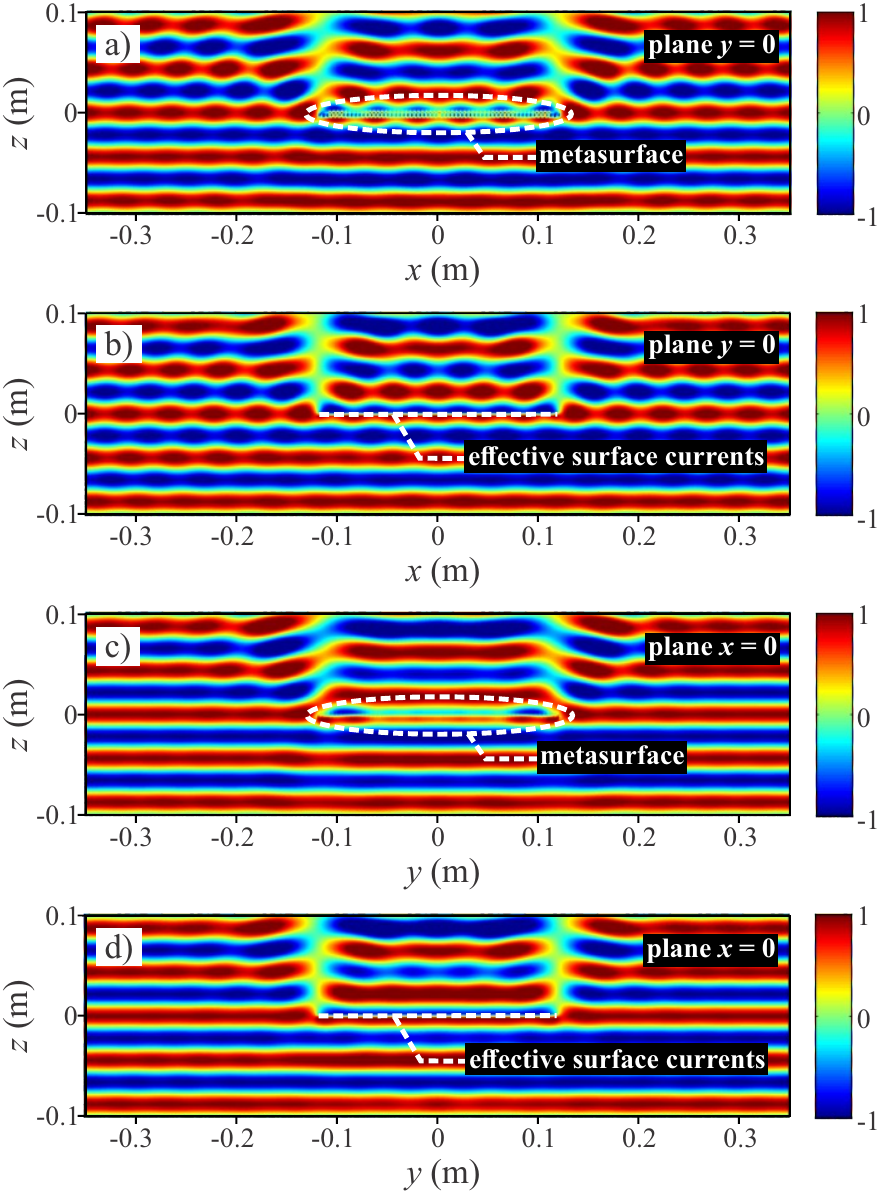} 
  \caption{Scattering by a finite MS at the resonance frequency ((a) and (c)) compared with the effect of the homogeneous effective surface currents ((b) and (d)). The incident plane wave is linearly polarized along the $y$-axis and propagating along the $z$-axis. Just the $y$-component of the total electric field normalized to the incident field is depicted on the central planes $y=0$ ((a) and (b)) and $x=0$ ((c) and (d)). The lateral size is $240$ mm ($\approx5$ resonance wavelengths) comprising $60\times60$ unit cells as the one shown in Fig.~\ref{Fig_sim_box} (with $s=2.4$~mm).}
  \label{Fig_sim_finite_sample}
\end{figure}

The particle dimensions considered were as depicted in Fig.~\ref{Fig_sim_box}; the particles were arranged in a $60 \times 60$ array for the finite sample, which resulted in an overall transverse dimension of $24~\text{cm} \times 24~\text{cm}$, just over five wavelengths per side at the resonant frequency. The excitation was provided by a plane wave propagating along $\hat z$ with $y$-directed electric field at the resonance frequency, which according to Fig.~\ref{Fig_sim_trans_refl}(c) will undergo the phase shift of 180\degree. The $y$-component of the total electric field was computed on $xz$- and $yz$-cuts around the surface in order to observe the transmitted wavefront, and the effect of reflection and diffraction from the MS. These results can be seen in Fig.~\ref{Fig_sim_finite_sample} (a) and (c), and demonstrate the expected phase shift in the portion right above the surface when compared to the reference wavefronts seen in areas far from the influence of the metasurface, where the original excitation wavefronts are virtually undisturbed; also the wavefront below the MS remains mostly unchanged both in phase and amplitude, indicating low reflection. Finally, the diffraction due to the finite sample size is visible mainly in the $xz$- cut (see Fig.~\ref{Fig_sim_finite_sample} a)).

In order to put in a better perspective the performance of our finite sample, it is instructive to analyze what can be achieved with an ideal scatterer with the same dimensions. The ideal scatterer in this case is a uniform continuous array of Huygens' sources. Given a normally-incident field with $\mathbf{E} = E_y \hat{\textbf{y}}$ and, the Huygens' source on the $xy$-plane that results in a phase jump of $180\degree$ is composed by electric and magnetic currents as follows: $\mathbf{J}_\text{s} = 2 Z_0^{-1} E_y \hat{\mathbf{y}}$, $\mathbf{K}_\text{s} = -2 E_y \hat{\mathbf{x}}$, where $Z_0$ is the characteristic impedance of free space. This ideal source was represented by the RWG basis on a triangular meshing of the $24~\text{cm} \times 24~\text{cm}$ aperture in order to compute the same field cuts as for the realistic MS before using AMDS. The RWG representation has the advantage of representing correctly the charge accumulation that takes place at the limits of the aperture, namely electric charge at the $\pm y$ limits and magnetic charge at the $\pm x$ limits. The total field is shown in Fig.~\ref{Fig_sim_finite_sample} (b) and (d). Interestingly, the behavior of this idealistic sheet is quite similar to that obtained with the finite realistic MS. In particular, similar diffraction contributions are seen, and these are stronger in the $xy$-plane, maintaining the spatial frequency. These results shows that the result obtained with the finite MS is very close to what can be expected from the idealized sample of the same size, and in some respects even better. Namely, the comparison of Fig.~\ref{Fig_sim_finite_sample} (c) and (d) shows a smoother wave front for the realistic MS, which results from the more natural tapering that takes place towards the edges of the the finite array (in the Huygens's source representation, current falls abruptly to zero at the aperture edges in the direction orthogonal to current flow). Reflection from these structures is also similar, as indicated by the almost undisturbed wavefronts for $z<0$ in any plot of Fig.~\ref{Fig_sim_finite_sample}.

\section{Experiment}\label{Experiment}
In order to verify the theoretical predictions, a realistic MS sample was manufactured and its transmission and reflection coefficients were measured in an anechoic chamber. The measurements were both under normal incidence with two orthogonal polarizations and under oblique incidence. Additional numerical simulations with CST Microwave Studio for an infinite MS corresponding to the experimental sample (with copper losses and a lossy substrate) were done to compare the computed transmission and reflection coefficients with the measured ones. These simulations were also done to adjust the geometric parameters of the elements in the unit-cell that satisfy the condition of decoupled SRRs and technological requirements.

\begin{figure}[h!]
  \centering
  \includegraphics[width=0.9\columnwidth]{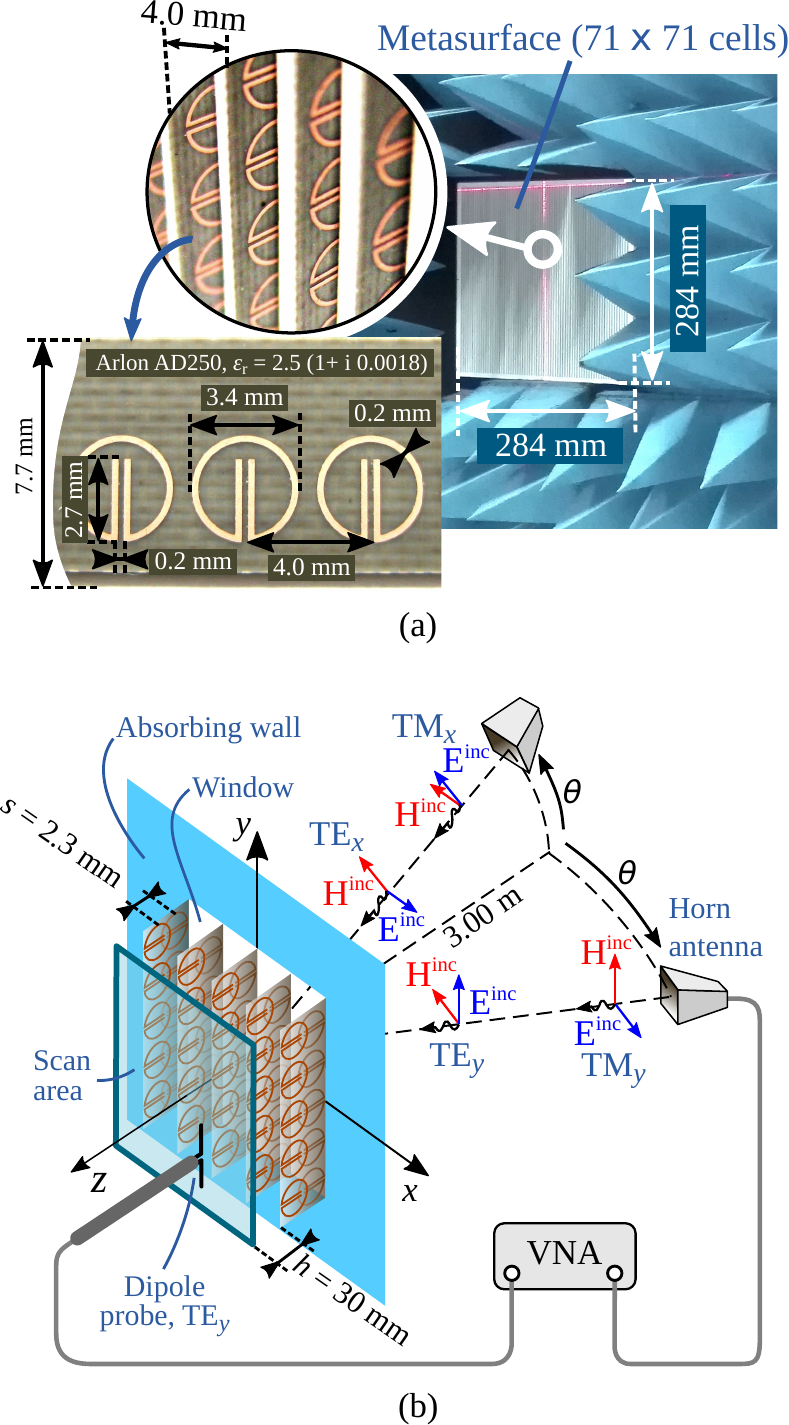} 
   \caption{Experimental setup. a) Photographs of the manufactured MS: the whole sample surrounded by microwave absorber, a few columns inserted into the foam holder, and a few resonators on one side of the PCB. b) Diagram of the experiment with the MS being illuminated with a horn antenna for different incidence angles and polarization states (TE$_x$, TE$_y$, TM$_x$ and TM$_y$). A dipole probe antenna was used for near-field scanning.The lateral shift between the patterns on both sides of the PCB is $s=2.3$~mm.}
   \label{Fig_setup}
\end{figure}

The MS in the experiment was composed of 71 identical Printed Circuit Board (PCB) columns arranged with the period $a=4$ mm as shown in the inset of Fig.~\ref{Fig_setup}(a). Each PCB column had the dimensions of $7.65~\text{mm} \times 284~\text{mm}$ and contained 71 identical SRR pairs. The two SRRs in each unit cell were placed on the opposite sides of an Arlon AD250 substrate with thickness 0.5~mm and relative permittivity $\varepsilon_r = 2.5(1+\text{i}*0.0018)$. All geometrical dimensions are shown in Fig.~\ref{Fig_setup}(a). It is worth noting that, in contrast to the previous idealized MS with $w=g=0.05~\text{mm}$ and $l=3$ mm, now the experimental MS has $w=g=0.2~\text{mm}$ and $l=2.7~\text{mm}$, just to avoid a significant frequency shift due to presence of the dielectric substrate. Moreover, the lateral shift between the two SRRs of the unit cell was slightly modified from $s=2.4$~mm to $s=2.3$~mm to optimize the decoupling. This modification can be explained by the fact that the substrate affected the profile of the current distribution along the rings of the SRRs due to modified capacity of the legs, and consequently the value of $s$ canceling their inductive coupling. All the PCB columns were mounted in milled slits in a 5-mm-thick foam holder. The photograph of the assembled MS with the dimensions of $284~\text{mm} \times 284~\text{mm}$ surrounded by microwave absorbers is shown in Fig.~\ref{Fig_setup}(a).

The transmission coefficients of the MS for the normal incidence and for the oblique incidence with the angles $\theta=15\degree$ and $\theta=30\degree$ and different canonical polarizations (TE$_x$, TE$_y$, TM$_x$, and TM$_y$) were measured using the setup schematically depicted in Fig.~\ref{Fig_setup}(b). The MS was installed in a square window surrounded by microwave absorbers and was illuminated using a broadband TEM-horn antenna (Trim, 1-20 GHz) located 3~m away. The main beam direction of the horn was pointed towards the center of the MS holding an angle $\theta$ with respect to the normal of the MS as shown in Fig.~\ref{Fig_setup}(b). For the TE$_y$ (TM$_y$) polarization the incident electric (magnetic) at the MS was along the $y$-axis for any angle of incidence. Analogously, for the TE$_x$ (TM$_x$) the electric field was along the $x$-axis for any angle of incidence.

In order to extract the transmission coefficients, the far-field transmitted by the MS was measured using the near-field approach as follows. The near-field distribution of the total electric field ($E_y$ for TE$_y$ and TM$_x$, while $E_x$ for TE$_x$ and TE$_y$) was scanned by positioning a small dipole-antenna probe polarized along the corresponding axis within the aperture of the MS. The scan area was located $h=30$ mm away from the MS as shown in Fig.~\ref{Fig_setup}(b), which is one order of magnitude longer than the periodicity of the MS. The dipole probe and the illuminating horn antenna were connected to two calibrated ports of the same Vector Network Analyzer (VNA). The near-field patterns were obtained for all the considered incidence angles in the same scan area by changing the position of the illuminating horn antenna and its polarization state. By applying the spatial Fourier transform to the near-field patterns, the far-field complex magnitude in the center of the transmitted beam was calculated. The same far-field values in the absence of the MS were also measured as a reference. The transmission coefficient for each $\theta$ and polarization state was determined as the ratio between the computed complex far-field magnitudes with and without the MS. The reflection coefficient was measured only for the normal incidence using the S-parameters measured by the VNA with the MS, without it and with a metal sheet of the same dimensions replacing the MS in the window as the reference reflector with the known reflection coefficient of -1. To reduce the undesired reverberations between the horn and the MS the time gating technique was applied. 

\begin{figure}
  \centering
  \includegraphics[width=0.9\columnwidth]{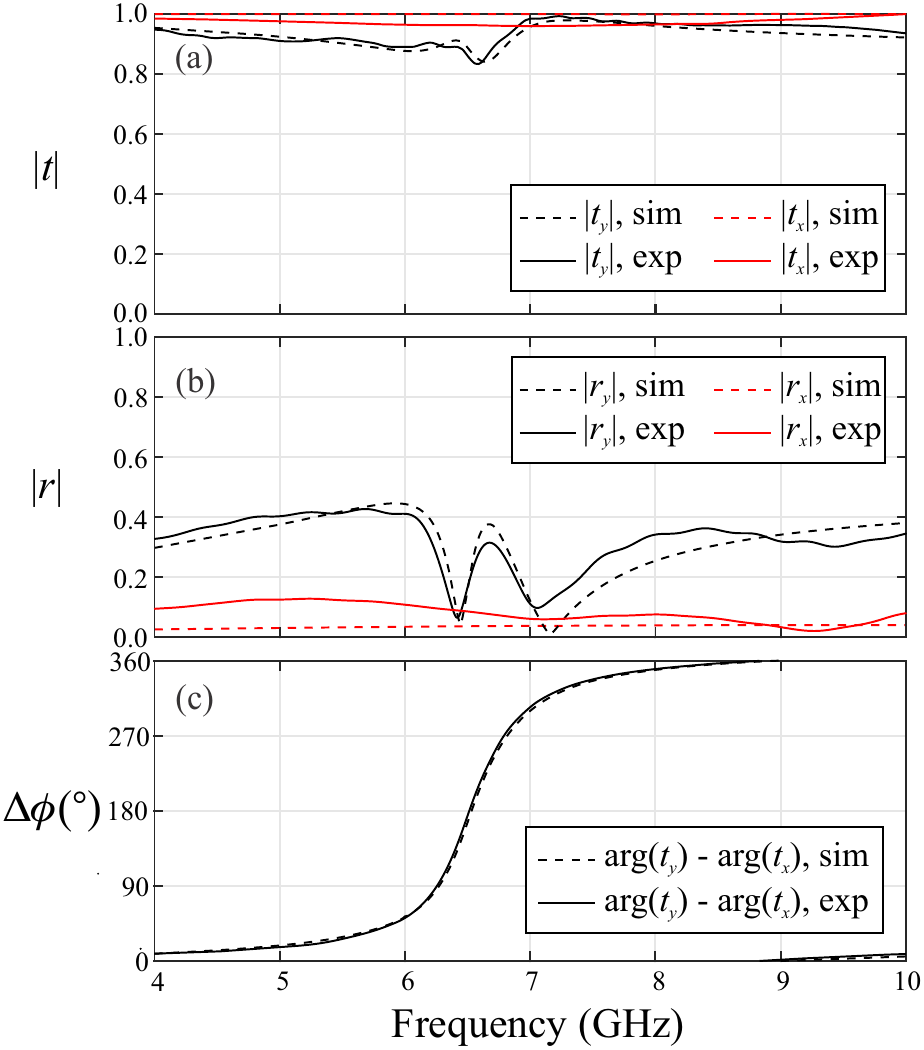}
  \caption{Measured and simulated transmission (a) and reflection (b) coefficient magnitudes under normal incidence as well as the phase jumps of the transmission coefficients (c).}
  \label{Fig_exp_normal}
\end{figure}

The transmission ($t_x$ and $t_y$) as well as the refection coefficients ($r_x$ and $r_y$) for the two orthogonal $x$- and $y$-polarizations measured under normal incidence are compared with simulation results in Fig.~\ref{Fig_exp_normal}. The transmission coefficients for both polarizations are presented in Fig.~\ref{Fig_exp_normal}(a). The measured $|t_y|$ is higher than 0.82 in the whole frequency range, i.e. the manufactured MS was indeed highly transparent. Also, it was almost transparent for the $x$-polarization as the measured $|t_x|$ was higher than 0.95. Moreover, as can be observed in Fig.~\ref{Fig_exp_normal}(b) the measured $|r_y|$ does not exceed 0.42. It is remarkable that, as follows from Fig.~\ref{Fig_exp_normal}(c), the difference between phases of the transmission coefficients for the $y$- and $x$-polarizations covers the whole range of $360\degree$ as the frequency changes around the resonance. Therefore, this MS is suitable for the full-range phase manipulation under the normal incidence keeping high transparency. The measured data is in a good agreement with the simulations.

\begin{figure}
  \centering
  \includegraphics[width=0.9\columnwidth]{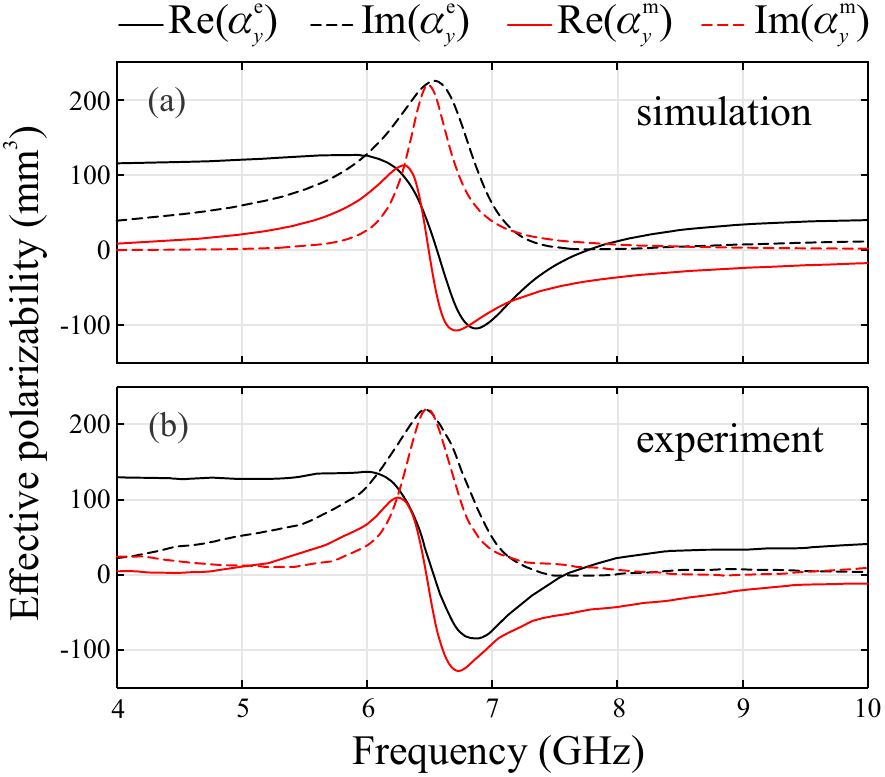}
  \caption{Electric and magnetic polarizabilities of the manufactured unit cell calculated with (\ref{boundaryConditions}) using simulated (a) and measured (b) transmission and reflection coefficients.}
  \label{Fig_exp_polarizabilities}
\end{figure}

Besides, to demonstrate the balanced hybridization of the electric and magnetic resonances, the electric and magnetic polarizabilities of the unit cell were retrieved from the transmission and reflection coefficients by using (\ref{boundaryConditions}). Results are depicted in Fig.~\ref{Fig_exp_polarizabilities} for simulated (a) and measured (b) polarizabilities, showing an excellent agreement. Although the electric and magnetic polarizabilities do not perfectly match one to another within the whole frequency range, they are very similar about the resonance frequency at 6.6~GHz.

\begin{figure}[h!]
  \centering
  \includegraphics[width=0.9\columnwidth]{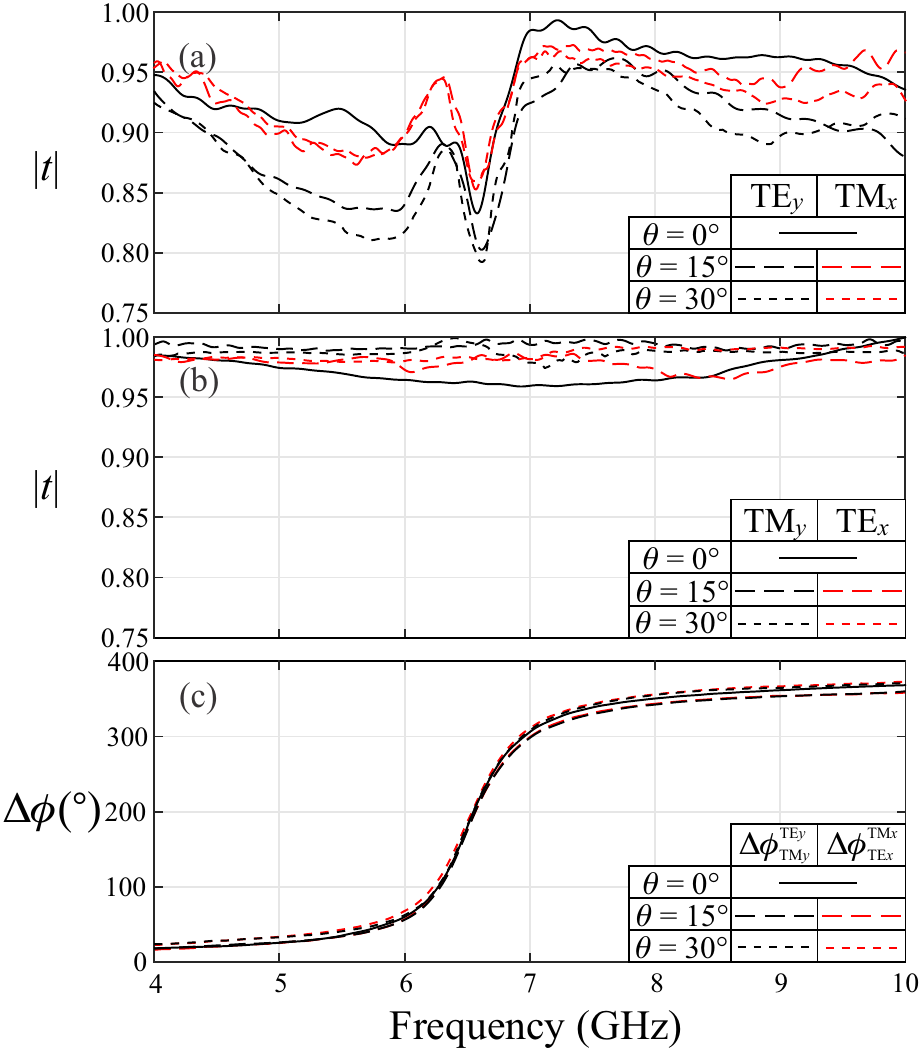}
\caption{Measured transmission coefficients under oblique incidence for different polarization states: TE$_y$ and TM$_x$ in (a), while TM$_y$ and TE$_x$ in (b). Also the phase differences for orthogonal polarizations corresponding to the same incidence directions is shown in (c).}
  \label{Fig_exp_oblique}
\end{figure}

To explore the angular stability of the MS response, the transmission coefficients were measured for off-normal incidence with $\theta=15\degree$ and $\theta=30\degree$, as shown in Fig.~\ref{Fig_setup}(b). The measured transmission coefficient magnitudes $|t|$ for TE$_y$ and TM$_x$ polarizations, shown in Fig.~\ref{Fig_exp_oblique}(a), demostrates high transmission with an absolute minimmum of $|t|=0.79$ at the resonance frequency (6.6~GHz). In Fig.~\ref{Fig_exp_oblique}(b) $|t|$ is given for TM$_y$ and TE$_x$ polarizations. In this case, $|t|$ is very flat and have an absolute minimum of $|t|=0.96$, i.e. the waves pass through the MS with very low interaction. Figures~\ref{Fig_exp_oblique}(a) and 11(b) also compare the off-normal incidence with the normal incidence. In general, curves for TE$_y$ and TM$_x$ looks similar to $|t_y|$ under normal incidence, while the cases of TE$_x$ and TM$_y$ are similar to $|t_x|$ under normal incidence. Finally, the transmission coefficient phases of waves with orthogonal polarizations propagating in the same direction of propagation are compared in Fig.~\ref{Fig_exp_oblique}(c) by plotting their difference (TE$_y$ with TM$_y$ and TE$_x$ with TM$_x$). In summary, for incidence angles up to $\theta=30\degree$, the MS exhibit high transparency ($|t|>0.79$ in the whole frequency range) and, at the same time, the phase differences cover the whole range from $0\degree$ to $360\degree$.

\section{Applications}
\label{Applications}
Metasurfaces have a number of applications that have been presented in the literature \cite{Glybovski2016}, and the design presented here provides specific advantages in view of its simple geometry, amenable for flush construction using the PCB technology that in turn affords low tolerances, mechanical robustness and high integrability in nowadays systems. Moreover, the design avoids combining two different types of unit cells, i.e. miniaturized dipoles and miniaturized loops, that need to be tuned separately. Some of the applications that can be envisioned for the design presented here are:
\begin{enumerate}

\item Arbitrary phase control: suitable design of the particles allows achieving arbitrary phase shifts with negligible reflection and even with dynamic control by means of varactors. By combining orthogonal rows of SRR pairs the design can be extended to an isotropic MS.

\item Arbitrary refraction: spatial modulation of the transmission coefficient's phase (by local geometric control of the particles) allows redirecting incident waves in any desired direction into the opposite hemisphere. Again, via the use of tuning elements such as varactors, dynamic control of the refraction angle can be achieved.

\item Polarization conversion: conversion between polarization states is achieved by a suitable orientation (rotation of the MS around z) and phase shift: helicity inversion is achieved by the MS in resonance, circular polarization turns into linear when the MS causes a phase shift of $\pm \pi/2~\mathrm{rad}$, and linear polarization is rotated 90 deg. with the MS in resonance and its y-axis rotated $\pm \pi/4 ~\mathrm{rad}$ with respect to the initial polarization.

\item Absorber: suitable lossy elements can be inserted such that the MS is still reflectionless, yet negligible energy is transmitted into the forward direction.

\item Matching interface: a partially reflective MS could be devised to provide an interface between media with disparate constitutive parameters such that total transmission is achieved.

\end{enumerate}

\section{Conclusions}

The phenomenon of broadband transparency with full-range phase control has been demonstrated from theory and simulation to an experiment by using Huygens' metasurface meta-atoms of a novel type each one containing two decoupled SRRs. It was demonstrated that being decoupled by a proper mutual lateral shift, an SRR pair possesses balanced electric and magnetic resonances at the same frequency.  This provided that the transmission coefficient magnitude is close to 1, while its phase can be controlled in a full 360\degree range. A new circuit model of the meta-atom has been developed and used for geometry optimization. 

The expected properties of the MS were validated by numerical simulations in the microwave range 0~GHz-10~GHz, with the resonance about 6.6~GHz. These numerical simulations with idealized substrate-less perfectly conducting SRRs demonstrated a transmission coefficient above 0.95 over the whole frequency band. For the realistic structure with copper SRRs etched on a dielectric substrate, the experimentally achieved transmission coefficient exceeded 0.8 in the range from 4 to 9~GHz. Moreover, the phase shift between two orthogonal polarization has been shown to cover the whole interval from $0\degree$ to $360\degree$ for incidence angles up to $30\degree$.

The achieved phase-shifting operation of the proposed metasurface with broadband transparency with respect to a selected linear polarization of an incident wave can be used for wavefront shaping in transmission. Moreover, we envisioned applications to narrowband conversion from linear to circular polarization or the orthogonal linear polarization at a certain angle of incident polarization. On the other hand, an isotropic version of the proposed MS could be constructed using two mutually orthogonal sets of pairs of SRRs.

Finally, we would like to point out that the tuning of the proposed Huygens' MS is much simpler than other previously reported geometries whose realizations are based on combinations of two different particles, electrically and magnetically polarizable. 

\section{Acknowledgments}
This research has been supported by Universidad Nacional de Colombia (Project No. HERMES-31154). The experimental work was financially supported by the Government of the Russian Federation (Grant 08-08).

\section*{Appendix: Effective electric dipole length and magnetic dipole area}

\begin{figure}
\centering
\includegraphics[width=0.9\columnwidth]{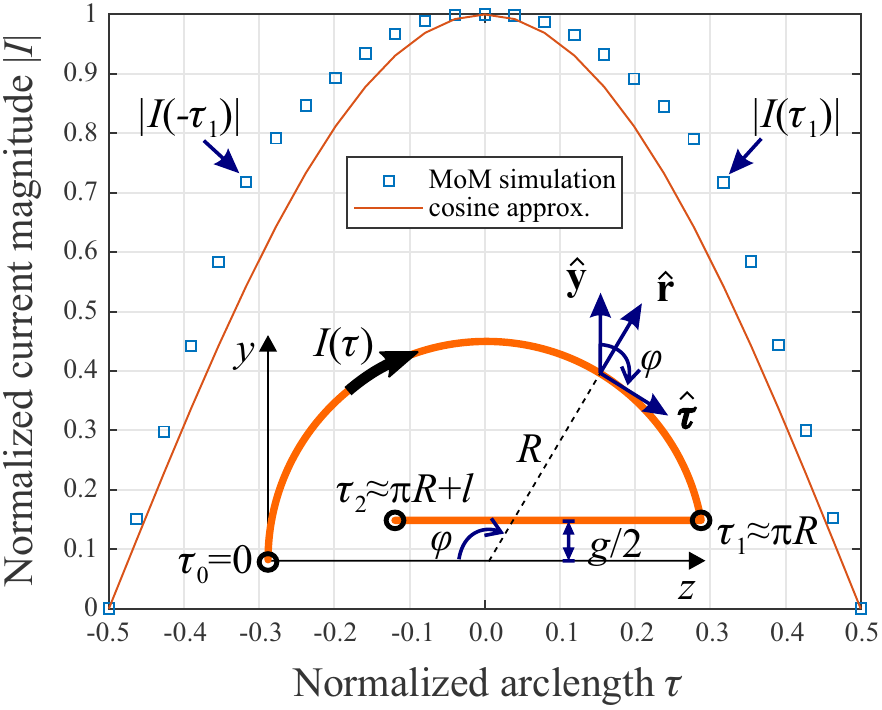}
\caption{Current distribution along one SRR obtained by using MoM at the resonance frequency. The inset shows a threadlike model of the SRR. The actual simulated geometry corresponds with the one shown in Fig.~\ref{Fig_sim_box} (with $s=2.4$~mm).}
\label{SRR}
\end{figure}

This appendix is devoted to the calculation of the effective electric dipole length and the effective magnetic dipole area of the used SRR. Let us assume the SRR is being illuminated by a $y$-polarized incident wave propagating along the $z$-axis. A threadlike model of one SRR is shown in the inset of Fig.~\ref{SRR} where the values of the arclength parameter $\tau$ are indicated for some cardinal points. To understand how the current is distributed along one SRR, we plotted the magnitude of the numerical current over different SRRs of the finite MS made of $60 \times 60$ unit cells, as reported in Section~\ref{FiniteSimulation}, and we found that all SRRs present approximately the same current pattern which is similar to a cosine function with nodes only at the ends. As an example, the current over a specific SRR near the center of the simulated finite MS is plotted in Fig.~\ref{SRR} and compared with the cosine function. 

In the harmonic regime, the electric dipole can be calculated by integrating the current density $\textbf{J}$ over the volume as $\textbf{p}=(i/\omega) \int \textbf{J} dv$. Due to the mirror symmetry, it is clear that the electric dipole is along the $y$-axis and the horizontal straight legs do not contribute. The $y$-component of the electric dipole can be approximatted as follows:
\begin{equation}
\label{electric_dipole}
\begin{gathered}
p_{y}= \frac{1}{j\omega}\int_{-\tau_2}^{\tau_2} I(\tau) \hat{\boldsymbol{\tau}} \cdot \hat{\textbf{y}} d\tau
\\
\approx \frac{2I(0)}{j\omega}\int_{0}^{\tau_1}\cos\left(\frac{\pi \tau}{2\tau_2}\right)\cos\left(\frac{\tau}{R}\right)d\tau 
\\
=\frac{I(0)}{j\omega}\left[2R\frac{sin(\pi\frac{\pi R}{2\pi R+2l})}{\frac{2\pi R+2l}{\pi R}-\frac{\pi R}{2\pi R+2l}}\right]
\end{gathered}
\end{equation}
Taking into account that the electric dipole of two point charges is defined as charge times distance, the term within the brackets could be considered as the distance between the positive and negative charge centers of the SRR. In fact, it coincides with the formula of the effective distance $d$ in (\ref{effGeoPar}).

On the other hand, the magnetic dipole is defined as $\textbf{m} = \frac{1}{2}\int_{V}\textbf{r}\times\textbf{J}dv$ and the only non-null component can be easily calculated as follows:
\begin{equation}
\label{magnetic_dipole}
\begin{gathered}
m_{x}=\frac{1}{2}\int_{-\tau_2}^{\tau_2} \hat{\textbf{x}} \cdot (\hat{\textbf{r}}\times\hat{\boldsymbol{\tau}}) R I(\tau) d\tau
\\
= R I(0) \int_{0}^{\tau_2} \cos\left(\frac{\pi \tau}{2\tau_2}\right) d\tau
\\
= I(0) \left[
  \pi R^{2}\frac{\sin\left(\pi\frac{\pi R}{2\pi R+2l}\right)}{\pi\frac{\pi R}{2\pi R+2l}}
\right]
\end{gathered}
\end{equation}
where the term within the brackets can be interpreted as the area of an effective magnetic loop supporting a uniform current $I(0)$. Thus, it coincides with the effective area $A$ mentioned in (\ref{effGeoPar}). 

\bibliography{bibliography}

\end{document}